\documentstyle[emulateapj,danonecolfloat]{article}

\begin{document}
\twocolumn[
\newcommand{\lya}{Lyman~$\alpha$}
\newcommand{\lyb}{Lyman~$\beta$}
\newcommand{\za}{$z_{\rm abs}$}
\newcommand{\ze}{$z_{\rm em}$}
\newcommand{\cmtwo}{cm$^{-2}$}
\newcommand{\nhi}{$N$(H$^0$)}
\newcommand{\degpoint}{\mbox{$^\circ\mskip-7.0mu.\,$}}
\newcommand{\kms}{\,km~s$^{-1}$}      
\newcommand{\minpoint}{\mbox{$'\mskip-4.7mu.\mskip0.8mu$}}
\newcommand{\peryr}{\mbox{$\>\rm yr^{-1}$}}
\newcommand{\secpoint}{\mbox{$''\mskip-7.6mu.\,$}}
\newcommand{\sqdeg}{\mbox{${\rm deg}^2$}}
\newcommand{\squig}{\sim\!\!}
\newcommand{\subsun}{\mbox{$_{\twelvesy\odot}$}}
\newcommand{\et}{{\rm et al.}~}

\def\ltsima{$\; \buildrel < \over \sim \;$}
\def\simlt{\lower.5ex\hbox{\ltsima}}
\def\gtsima{$\; \buildrel > \over \sim \;$}
\def\simgt{\lower.5ex\hbox{\gtsima}}
\def\arcs{$''~$}
\def\arcm{$'~$}
\def\erf{\mathop{\rm erf}}
\def\erfc{\mathop{\rm erfc}}

\title{X-Ray and Radio Emission from UV-Selected Star Forming Galaxies 
at Redshifts $1.5 \la Z \la 3.0$ in the GOODS-North Field\altaffilmark{1}}
\author{\sc Naveen A. Reddy \& Charles C. Steidel}
\affil{California Institute of Technology, MS 105--24, 
Pasadena, CA 91125; nar@astro.caltech.edu, ccs@astro.caltech.edu}

\slugcomment{DRAFT: \today}

\altaffiltext{1}{Based on data obtained at the W. M. Keck Observatory, which is
operated as a scientific partnership among the California Institute of Technology,
the University of California, and NASA, and was made possible by the generous financial
support of the W. M. Keck Foundation.}

\begin{abstract}

We have examined the stacked radio and X-ray emission from UV-selected galaxies spectroscopically
confirmed to lie between redshifts $1.5\la z\la 3.0$ in the GOODS-North field
to determine their average extinction and star formation rates (SFRs).  The X-ray and radio 
data are obtained from the Chandra 2~Msec survey and the Very Large Array, respectively.  
There is a good agreement between the X-ray, radio, and de-reddened UV estimates of the average 
SFR for our sample of $z\sim2$ galaxies of $\sim 50$~M$_\odot$~yr$^{-1}$, indicating that
the locally-calibrated SFR relations appear to be statistically valid from redshifts $1.5\la z\la 3.0$.
We find that UV-estimated SFRs (uncorrected for extinction) underestimate the bolometric SFRs 
as determined from the $2-10$~keV X-ray luminosity by a factor of $\sim 4.5 - 5.0$ for galaxies 
over a large range in redshift from $1.0\la z\la 3.5$.  

\end{abstract}

]

\section{Introduction}

Estimating global star formation rates (SFRs) of galaxies typically requires using 
relations that can be quite uncertain as they incorporate a large number of 
assumptions in converting between specific and bolometric luminosities 
(e.g., assumed IMF, extinction, etc.; e.g., Adelberger \& Steidel 2000).  The varied 
efforts in the Great Observatories Origins Deep Survey (GOODS; Giavalisco et~al. 2003) 
allow us to examine the same galaxies over a broad range of wavelengths to mitigate 
some of these uncertainties.  X-ray, radio, and UV emission are all thought to 
directly result from massive stars and are consequently used as tracers of current
star formation (e.g., Ranalli, Comastri, \& Setti 2003; Condon 1992; Gallego et~al. 1995).  
Here we use the X-ray, radio, and UV emission,
each differently affected by extinction (or not at all), to determine SFRs 
of galaxies at $z\sim 2$.

Observations of the QSO and stellar mass density, and morphological diversification
all point to the epoch around $z\sim 2$ as an important period in cosmic history (e.g.,
Di Matteo et~al. 2003; Chapman et~al. 2003).
Until recently, this epoch has been largely unexplored as lines used for redshift 
identification are shifted to the near-UV where detector sensitivity has been poor or to
the near-IR, where spectroscopy is more difficult due to higher backgrounds.
With the recent commissioning of the blue side of the Low Resolution Imaging Spectrograph
(LRIS; Oke et~al. 1995) on the Keck I telescope, we have for the first time been able to 
obtain spectra for large numbers of galaxies at these redshifts.  
Adding to the multi-wavelength efforts 
in the GOODS-North field, we have undertaken a program to identify photometric candidates in this 
field between $1.5\la z\la 3.0$ and perform followup spectroscopy with LRIS-B 
(Steidel et~al. 2004).  This UV-selected sample of galaxies forms
the basis for our subsequent multi-wavelength analysis.

Current sensitivity limits at X-ray and radio wavelengths preclude the 
direct detection of normal star forming galaxies at $z\ga 1.5$.  Nonetheless, we can use
a ``stacking'' procedure to effectively add the emission from a class of objects in order
to determine their average emission properties (e.g., Nandra et~al. 2002; Brandt et~al. 2001;
Seibert, Heckman, \& Meurer 2002).  In this paper, we present a stacking
analysis of the radio and X-ray emission from UV-selected star forming galaxies at redshifts 
$1.5\la z\la 3.0$ to cross-check three different techniques of estimating SFRs at high
redshifts.
$H_{o}=70$~km~s$^{-1}$~Mpc$^{-1}$, $\Omega_{\rm M}=0.3$, and
$\Omega_{\Lambda}=0.7$ are assumed throughout.

\section{Data}

The techniques for selecting galaxies at $z\sim 2$ are designed to cover the same range of
UV properties and extinction to those used to select Lyman-break galaxies (LBGs) at higher 
redshifts ($z\ga 3.0$; Adelberger et~al. 2004).  
Here, we simply mention that we have two spectroscopic samples at $1.5\la z\la 2.5$: 
a ``BX'' sample of galaxies selected on the expected $U_{\rm n}G\cal R$ colors 
of LBGs de-redshifted to $2.0\la z\la 2.5$; and a ``BM'' sample 
targeting $z\sim 1.5-2.0$.  (see Adelberger et~al. 2004 and Steidel et.~al. 2004 for a complete
description).  We presently have 138 redshifts ($\langle z\rangle \sim 2.2\pm 0.3$) 
and 48 redshifts ($\langle z\rangle \sim 1.7\pm 0.3$) in the GOODS-North BX and BM samples, 
respectively.

The X-ray data are from the Chandra 2~Msec survey of the GOODS-North region 
(Alexander et~al. 2003).  We made use of their raw images 
and exposure maps in the Chandra soft X-ray band ($0.5-2.0$~keV).
Dividing the raw image by the appropriate exposure map yields an image
with the count rates corrected for vignetting, exposure time, and variations in
instrumental sensitivity.  The on-axis soft band sensitivity is 
$\sim 2.5\times 10^{-17}$~ergs~cm$^{-2}$~s$^{-1}$ (3~$\sigma$).

The radio data are from the Richards (2000) Very Large Array (VLA) survey 
of the Hubble Deep Field North (HDF-N), reaching a 3~$\sigma$ sensitivity of 
$\sim 23\, \mu$Jy~beam$^{-1}$ 
at 1.4~GHz.  The final naturally-weighted image has a pixel size of $0\farcs 4$ and 
resolution of $2\farcs 0$, with astrometric accuracy of $< 0\farcs 03$.

\section{Stacking Procedure}

We divided the spectroscopic data into subsets based on
selection (BX and BM) and redshift, removed
sources with matching X-ray or radio counterparts within $2\farcs 5$ (or sources
whose apertures are large enough to contain emission from a nearby extended X-ray 
or radio source), and stacked galaxies in these subsets.  Four of the removed x-ray/radio
sources are detected at $850$~${\rm \mu m}$ with SCUBA.

The X-ray data were stacked using the following procedure.  We added the flux 
within apertures randomly dithered by $0\farcs 5$ at the positions of the galaxies 
(targets) in the X-ray images to produce a signal.  
Similarly-sized apertures were randomly placed within $5\arcsec$ of the galaxy positions to 
sample the local background near each galaxy while avoiding known X-ray
sources.  This was repeated 1000 times to estimate the mean signal and
background.  The Chandra PSF widens for large angles from the average 
pointing (off-axis angle), and we fixed the aperture sizes to the $50\%$ 
encircled energy (EE) radii (Feigelson et~al. 2002) for off-axis angles $> 6'$.  
Background included at large off-axis angles
becomes significant due to increasing aperture size and this can degrade the total 
stacked signal.  Consequently, we only stacked galaxies within the off-axis angle 
that results in the highest S/N (this varies for each subsample, from $6$ to $8'$).
Including all sources in the stack reduces the S/N but does not affect the absolute
flux value.  For sources $<6'$ from the pointing center, the $50\%$ EE radius falls 
below $2\farcs 5$, and we adopted a fixed $2\farcs 5$ radius aperture to avoid the 
possibility of placing an aperture off a target as a result 
of dithering or astrometric errors---which are $\sim 0\farcs 3$---for sources very close to the 
average pointing.  Stacking was performed on both the raw and normalized images to
calculate the S/N and total count rate, respectively.
Aperture corrections were applied to the raw counts and count rate.
The conversion between count rate and flux was determined by averaging the count rate
to flux conversions for the $74$ optically bright X-ray sources in the catalogs of 
Alexander et~al. (2003; Table~7) which are assumed to have a photon index of $\Gamma = 2.0$,
typical of the X-ray emission from star forming regions (e.g., Kim, Fabbiano, \& Trinchieri
1992; Nandra et~al. 2002), 
and incorporate corrections for the QE degredation of the ACIS-I chips.  In converting
flux to rest-frame luminosity, we assume $\Gamma = 2.0$ and a Galactic absorption column
density of $N_{\rm H} = 1.6\times 10^{20}$~cm$^{-2}$ (Stark et~al. 1992).  Uncertainties
in flux and luminosity are dominated by Poisson noise and not the dispersion in measured
values for each stacking repetition, so we assume the former.  

To stack the radio data, we extracted sub-images at the locations of the targets from the 
mosaicked radio data of Richards (2000).  These were corrected for the primary beam
attenuation of the VLA with a maximum gain correction of $15\%$, coadded using a
$1/\sigma^2$ weighted average to produce a stacked signal with maximal $S/N\sim4.5$, and 
smoothed by $1\farcs5$.  The integrated flux density, $S_{\rm 1.4\, GHz}$, and error were 
computed from the standard AIPS\footnote{AIPS is the Astronomical Image Processing System 
software package written and supported by the National Radio Astronomy Observatory.} task 
JMFIT using an elliptical gaussian to model the stacked emission.  We assume a synchrotron 
spectral index of $\gamma = -0.8$, typical of the non-thermal radio emission from star 
forming galaxies (Condon 1992).  Results of the X-ray and radio stacks for various subsamples are 
presented in Table~\ref{tab:stackres}.  Four subsamples contain too few sources 
to yield a robust estimate of the stacked radio flux.

\begin{deluxetable}{lcccccccc}
\tablewidth{0pc}
\tablecaption{Radio and X-ray Stacking Results}
\tablehead{
\colhead{} &
\colhead{} &
\colhead{} &
\colhead{} &
\colhead{$F_{\rm 0.5-2.0\, keV}$\tablenotemark{d}} &
\colhead{$L_{\rm 2.0-10\, keV}$\tablenotemark{e}} &
\colhead{$S_{\rm 1.4\, GHz}$\tablenotemark{f}} &
\colhead{$L_{\rm 1.4\, GHz}$\tablenotemark{g}} &
\colhead{$\nu L_{\nu}$\tablenotemark{h}} \\
\colhead{Sample} & 
\colhead{$N_{\rm s}$\tablenotemark{a}} &
\colhead{$\langle z\rangle$\tablenotemark{b}} &
\colhead{S/N\tablenotemark{c}} &
\colhead{($\times 10^{-18}$~ergs~cm$^{-2}$~s$^{-1}$)} &
\colhead{($\times 10^{41}$~ergs~s$^{-1}$)} &
\colhead{($\mu$Jy)} &
\colhead{($\times 10^{22}$~W~Hz$^{-1}$)} &
\colhead{($\times 10^{10}$~L$_{\odot}$)}}
\startdata
BX+BM & 147 & 2.09 & 8.9 & $5.65\pm0.68$ & $2.11\pm0.25$ & $2.30\pm0.65$ & $5.90\pm1.66$ & 3.50\\
BX & 109 & 2.22 & 6.8 & $4.83\pm0.79$ & $2.09\pm0.34$ & $2.09\pm0.75$ & $6.17\pm2.21$ & 3.86\\
BM & 38 & 1.71 & 6.0 & $8.04\pm1.34$ & $1.84\pm0.31$ & ... & ... & 2.46\\
$1.5<z\leq2.0$ & 54 & 1.82 & 5.6 & $6.89\pm1.27$ & $1.84\pm0.33$ & ... & ... & 2.81\\
$2.0<z\leq2.5$ & 73 & 2.24 & 6.0 & $5.24\pm0.96$ & $2.33\pm0.43$ & ... & ... & 4.05\\
$2.5<z\leq3.0$ & 43 & 2.87 & 3.3 & $4.21\pm1.46$ & $3.40\pm1.18$ & ... & ... & 4.61\\
\enddata
\tablenotetext{a}{Number of galaxies stacked}
\tablenotetext{b}{Mean redshift}
\tablenotetext{c}{Signal-to-noise calculated in a manner analogous to that
in Nandra et~al. 2002}
\tablenotetext{d}{Average soft-band X-ray flux per object}
\tablenotetext{e}{Average rest-frame X-ray luminosity per object, assuming $\Gamma = 2.0$ and 
$N_{\rm H} = 1.6\times 10^{20}$~cm$^{-2}$, for our adopted cosmology}
\tablenotetext{f}{Average integrated radio flux density per object}
\tablenotetext{g}{Average rest-frame 1.4~GHz luminosity per object, assuming synchrotron
spectral index of $\gamma = -0.8$}
\tablenotetext{h}{Average UV luminosity computed from $G$ and $\cal R$ magnitudes approximating
the $1600$ and $1800$~${\rm \AA}$ fluxes, respectively.}
\label{tab:stackres}
\end{deluxetable}

\section{Results and Discussion}

\subsection{SFR Estimates}

The relations established at $z=0$ to convert luminosities to 
SFRs for our $z\sim 2$ sample are adopted from the following sources:  Kennicutt (1998a,b)
for conversion of the $1500-2800$~${\rm \AA}$ luminosity; Ranalli et~al. (2003)
for the $2-10$~keV luminosity; and Yun, Reddy, \& Condon (2001) for 
the $1.4$~GHz luminosity.  These relations must be used with caution when applied to 
individual galaxies 
given uncertainties in the SFR relations (e.g., burst age, IMF) as well as the factor of $\sim 2$ 
dispersion in the correlations between different specific luminosities.  However, they
should yield reasonable results when applied to an ensemble of galaxies,
as we have done here.

Table~\ref{tab:sfrs} shows the SFR estimates based on the $2-10$~keV (``SFR$_{\rm X}$''),
$1.4$~GHz (``SFR$_{\rm 1.4~GHz}$''), and UV (``SFR$_{\rm UV}$'') luminosities, with typical
error of $\sim 20\%$. 
We approximate the UV luminosity by using the 
$1600$~${\rm \AA}$ rest-frame flux for all samples except the highest
redshift bin sample ($2.5<z\la 3.0$) where we use the $1800$~${\rm \AA}$
rest-frame flux.  UV-estimated SFRs were corrected for extinction using the 
observed $G-\cal R$ colors, a spectral template assuming constant star formation for 
$>10^8$~yr (after which the UV colors are essentially constant), and applying the reddening 
law of Calzetti et~al. (2000) and Meurer, Heckman, \& Calzetti (1999).  We created
4 additional subsamples of galaxies according to de-reddened UV-estimated SFR, also
shown in Table~\ref{tab:sfrs}.

\begin{deluxetable}{lcccc}
\tablewidth{0pc}
\tablecaption{Star Formation Rate Estimates}
\tablehead{
\colhead{} &
\colhead{SFR$_{\rm X}$} &
\colhead{SFR$_{\rm R}$} &
\colhead{SFR$_{\rm UV}^{cor}$} &
\colhead{} \\
\colhead{Sample} &
\colhead{(M$_{\odot}$~yr$^{-1}$)} &
\colhead{(M$_{\odot}$~yr$^{-1}$)} &
\colhead{(M$_{\odot}$~yr$^{-1}$)} &
\colhead{SFR$_{\rm X}$/SFR$_{\rm UV}^{uncor}$}}
\startdata
BX+BM & 42 & 56 & 50 & 4.5 \\
BX & 42 & 58 & 54 & 4.2 \\
BM & 37 & ... & 38 & 4.8 \\
$1.5<z\leq2.0$ & 37 & ... & 49 & 4.3 \\
$2.0<z\leq2.5$ & 47 & ... & 57 & 4.4 \\
$2.5<z\leq3.0$ & 68 & ... & 70 & 4.7 \\
SFR$_{\rm UV}^{cor}\leq20$~M$_{\odot}$~yr$^{-1}$ & 14 & ... & 11 & 2.3 \\
$20<$SFR$_{\rm UV}^{cor}\leq40$~M$_{\odot}$~yr$^{-1}$ & 40 & ... & 38 & 4.4 \\
$40<$SFR$_{\rm UV}^{cor}\leq60$~M$_{\odot}$~yr$^{-1}$ & 44 & ... & 48 & 4.7 \\
SFR$_{\rm UV}^{cor}>60$~M$_{\odot}$~yr$^{-1}$ & 72 & ... & 73 & 5.1 \\
\enddata
\label{tab:sfrs}
\end{deluxetable}

\subsection{Stacked Galaxy Distribution and AGN}

\begin{figure}
\plotone{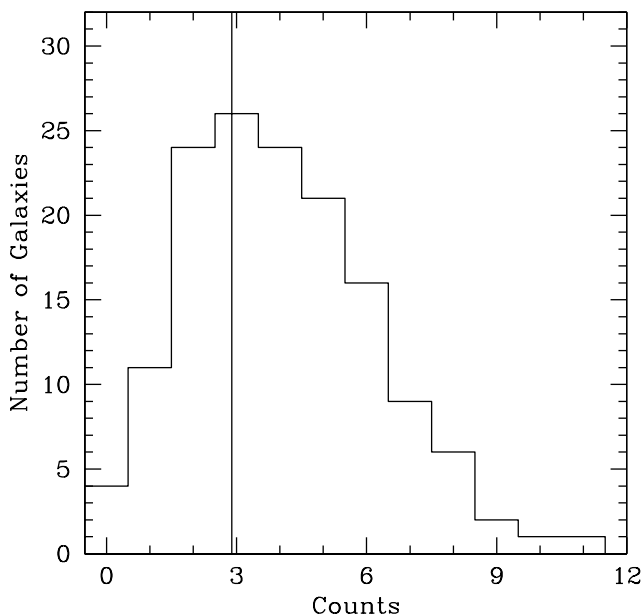}
\caption{Average distribution of counts for the spectroscopic sample.  The vertical line
denotes the average background count per aperture.  The number excess at high
counts ($>7$) results from random positive fluctuations.
\label{fig:cntdist}}
\end{figure}

Stacking only indicates the average emission properties of galaxies, not their actual 
distribution, and the observed signal may result from a few luminous 
sources lying just below the detection threshold.  To investigate this, 
we plot the average distribution in counts for the sample of 147 stacked spectroscopic galaxies 
(Figure~\ref{fig:cntdist}).  Much of the high-end tail of the distribution results from  
random positive fluctuations.  Only $3$ sources consistently had $>7$ counts.
Removing those objects whose apertures have $>6$ counts
still results in a stacked signal with $S/N \sim 2.5$ and an average loss of $21$ 
galaxies ($\sim 14\%$ of the sample).  It is therefore likely that most
of the stacked galaxies contribute to the signal, particularly given their wide 
range in optical, and likely X-ray, properties. 

Contribution of low luminosity AGNs to the stacked signal is a
concern.  This is a problem with most X-ray stacking analyses, but
we also possess the UV spectra for our sources.  There 
are two objects undetected in X-rays for which the UV spectra show emission lines 
consistent with an AGN.  Our ability to identify AGNs from their UV spectra regardless
of their X-ray properties, and having identified only 2 such objects out of 149,
suggests that sub-threshold luminous AGNs do not contribute significantly to the stacked
X-ray flux.  Furthermore, UV selection biases against the dustiest sources so we
do not expect to find many Compton-thick AGNs in our sample.

There are also two BM galaxies coincident with known radio sources that are not 
detected in X-rays and are not included in the stacked samples.  Removing such
objects ensures excluding radio-loud AGN that might have unassuming UV and X-ray properties.
For comparison, the $3~\sigma$ radio sensitivity is sufficient to detect 
SFR$\ga 170$~M$_\odot$~yr$^{-1}$, a factor of $4$ higher than the median SFR of our
sample based on the X-ray or de-reddened UV SFR estimates.  The stacked X-ray emission 
indicates a SFR of $\sim 42$~M$_\odot$~yr$^{-1}$.  The on-axis soft-band flux limit
implies a sensitivity to SFR$\ga 186$~M$_\odot$~yr$^{-1}$ at $z\sim 2$, a factor of 
$4.5$ higher than the average SFR for spectroscopic $z\sim 2$ galaxies.
Stacking the radio flux for the full spectroscopic sample indicates 
average SFRs from $33-70$~M$_\odot$~yr$^{-1}$ depending on which estimator is used:
the Bell (2003), Condon (1992), and Yun et~al. (2001) calibrations give
low, high, and median ($\sim 56$~M$_\odot$~yr$^{-1}$) values,
assuming $\gamma = -0.8$.  We adopted the Yun et~al. (2001) conversion (corrected for
a binning error) as it is most relevant to the radio luminosity range considered here.

\subsection{Bolometric Properties of $z\sim 2$ Galaxies}

The UV-implied reddening indicates $A_{\rm V}\sim0.5$~mag and $N_{\rm H} \sim 7.5\times 10^{20}$~cm$^{-2}$
assuming the Galactic calibration (Diplas \& Savage 1994).  For this column density, absorption
in the $2-10$~keV band is negligible, and we therefore assume that SFR$_{\rm X}$ is
indicative of the bolometric SFR.  In this case, we find a good agreement between the
SFRs determined from the X-ray, radio, and de-reddened UV luminosities ($L_{\rm X}$, $L_{\rm 1.4~GHz}$, 
and $L_{\rm UV}$), suggesting that the locally-calibrated relations between specific luminosity 
and SFR remain valid within the uncertainties at $z\sim 2$, under the caution that we cannot 
independently test for these relations as we have no direct measure of $L_{\rm bol}$.

The $\langle L_{\rm X}\rangle$ and $\langle L_{\rm 1.4~GHz}\rangle$ of spectroscopically 
identified $z\sim2$ galaxies are 
comparable to those of local starbursts.  The X-ray/FIR relation for local galaxies 
(Ranalli et~al. 2003) 
implies $\langle L_{\rm FIR}\rangle \sim 2.6\times10^{11}$~L$_{\odot}$.  
The stacked $L_{\rm 1.4~GHz}$ implies $\langle L_{\rm FIR}\rangle = 1.1\times10^{11}$~L$_{\odot}$ 
(Yun et~al. 2003).
These estimates are similar to the FIR luminosity of luminous infrared galaxies (LIRGs), and
are expected to have 
$S_{\rm 850\mu m} \sim 0.3$~mJy (e.g., Webb et~al. 2003) and would therefore be missing in 
confusion-limited SCUBA surveys to $2$~mJy.  
SIRTF will have the same rest-frame $7$~${\rm \mu m}$ sensitivity to $z\sim 2$ galaxies
as ISO has at $z\sim 1$ for $L_{\rm FIR} \ga 5\times 10^{10}$~L$_{\odot}$ galaxies 
(e.g., Weedman, Charmandaris, \& Zezas 2004; Flores et~al. 1999).  Therefore, unlike the stacked
averages presented here, the SIRTF data will be the first extinction-free tracer of the 
SFR distribution of the $z\sim 2$ sample as the stacked galaxies should be individually
detected at $24$~${\rm \mu m}$.

For a fair comparison between the three redshift bins for $1.5<z\la 3.0$, we have added back those 
direct X-ray detections in the stacks for the 
$1.5<z\le2.0$ and $2.0<z\le2.5$ samples that would not have been detected if they had $z>2.5$.
There were no such sources with $1.5<z\le2.0$ and only one with $2.0<z\le2.5$, increasing 
$\langle L_{\rm 2-10~keV}\rangle$ by $2\%$ to $2.38\times10^{41}$~ergs~s$^{-1}$.  

The distance independent ratio SFR$_{\rm X}/$SFR$_{\rm UV}^{uncor}$ (Table~\ref{tab:sfrs}) is similar 
among the selection and redshift subsamples indicating that on average {\it UV-estimated SFRs (uncorrected for
extinction) are a factor of $\sim 4.5$ times lower than the bolometric SFRs for galaxies between
redshifts $1.5<z\le3.0$}.  For comparison, Nandra et~al. (2002) find this factor to be
$\sim 5$ for both the $z\sim1$ BBG and $z\sim3$ LBG populations, and the factor is comparable
to that of local starburst galaxies (Seibert et~al. 2002).  
The attenuation computed for the BX/BM sample using the Calzetti et~al. (2000) 
extinction law is similar to that computed from SFR$_{\rm X}/$SFR$_{\rm UV}^{uncor}$.  
The de-reddened UV-estimated SFRs (SFR$_{\rm UV}^{cor}$) agree well with those 
predicted from the radio continuum for the two samples for which radio estimates could be obtained. 
Finally, we note the factor of $\sim 5$ UV attenuation is similar to that advocated
by Steidel et~al. (1999) for UV-selected samples at all redshifts.

The average attenuation factor increases as the SFR increases, as shown by the last 4 subsamples 
in Table~\ref{tab:sfrs}, and is expected if galaxies with higher SFRs have greater
dust content on average (e.g., Adelberger \& Steidel 2000).  SFR$_{\rm UV}^{cor}$ follows the 
bolometric SFR even for low luminosity 
systems, indicating that the observed correlations are not entirely driven by only the most luminous galaxies.

\section{Conclusions}

We have made significant progress in estimating and comparing SFRs determined 
from UV, X-ray, and radio emission from galaxies between redshifts $1.5\la z \la 3.0$, postulated 
to be the most ``active'' epoch for galaxy evolution.  The locally-calibrated SFR relations, though 
uncertain in individual systems, appear to remain statistically valid at high redshift.
Stacking the X-ray and radio emission from UV-selected galaxies at $z\sim 2$ indicates that these
galaxies have an average SFR of $\sim 50$~M$_\odot$~yr$^{-1}$ and an average UV attenuation factor
of $\sim 4.5$.  
The prospect of increased radio sensitivity with the E-VLA, as well
as X-ray campaigns in different fields to similar depth as the $2$~Msec survey in the GOODS-North
field,
will allow for a more direct probe of the radio and X-ray flux distribution for
the stacked galaxies.  SIRTF/MIPS $24$~${\rm \mu m}$ data for the GOODS-N field will
trace the dusty star formation in $z\sim 2$ galaxies and allow for the cross-checking
of the results presented here.

\acknowledgements
We thank Alice Shapley, Dawn Erb, Matt Hunt, and Kurt Adelberger
for help in obtaining the data presented here.  CCS has been supported 
by grants AST 0070773 and 0307263 from the National Science Foundation 
(NSF) and by the David and Lucile Packard Foundation.  NAR acknowledges 
support from a NSF Graduate Research Fellowship.


\begin{references}

\reference{} Adelberger, K.~L., \& Steidel, C.~C. 2000, ApJ, 544, 218
\reference{} Adelberger, K.~L., Steidel, C.~C., Shapley, A.~E., Hunt, M.~P., Erb, D.~K., 
             Reddy, N.~A., \& Pettini, M. 2004, submitted to ApJ
\reference{} Alexander, D.~M., Bauer, F.~E., Brandt, W.~N., Schneider, D.~P., Hornschemeier,
             A.~E., Vignali, C., Barger, A.~J., Broos, P.~S., et al. 2003, AJ, 126, 539
\reference{} Bell, E.~F. 2003, ApJ, 586: 794
\reference{} Brandt, W.~N., Hornschemeier, A.~E., Schneider, D.~P., Alexander, D.~M., 
             Bauer, F.~E., Garmire, G.~P., \& Vignali, C. 2001, ApJ, 558, 5
\reference{} Calzetti, D., Armus, L., Bohlin, R.~C., Kinney, A.~L., Koornneef, J., \&
             Storchi-Bergmann, T. 2000, ApJ, 533, 682
\reference{} Chapman, S.~C., Blain, A.~W., Ivison, R.~J., \& Smail, I.~R. 2003, Nature, 422, 695
\reference{} Condon, J.~J. 1992, ARA\&A, 30, 575
\reference{} Di Matteo, T., Croft, R.~A.~C., Springel, V., Hernquist, L. 2003, ApJ, 593, 56
\reference{} Diplas, A. \& Savage, B.~D. 1994, ApJS, 93, 211
\reference{} Feigelson, E.~D., Broos, P.~S., Gaffney, J.~A. III, Garmire, G., Hillenbrand,
             L.~A., Pravdo, S.~H., Townsley, L., \& Tsuboi, Y. 2002, ApJ, 574, 258
\reference{} Flores, H., Hammer, F., Thuan, T.~X., C\'{e}sarsky, C., Desert, F.~X., Omont, A., 
             Lilly, S.~J., Eales, S., et~al. 1999, ApJ, 517, 148
\reference{} Gallego, J., Zamorano, J., Arag'on-Salamanea, A., \& Rego, M. 1995, ApJ, 455, 1
\reference{} Giavalisco, M., Ferguson, H.~C., Koekemoer, A.~M., Dickinson, M., Alexander, D.~M.,
             Bauer, F.~E., Bergeron, J., Biagetti, C., et~al. 2004, ApJL, 600, 93
\reference{} Kennicutt, R.~C. 1998a, ARA\&A, 36, 189
\reference{} Kennicutt, R.~C. 1998b, ApJ, 498, 541
\reference{} Kim, D.-W., Fabbiano, G., \& Trinchieri, G. 1992, ApJ, 393, 134
\reference{} Meurer, G.~R., Heckman, T.~M., \& Calzetti, D. 1999, ApJ, 521, 64
\reference{} Nandra, K., Mushotzky, R.~F., Arnaud, K., Steidel, C.~C., Adelberger, K.~L.,
             Gardner, J.~P., Teplitz, H.~I., \& Windhorst, R.~A. 2002, ApJ, 576, 625
\reference{} Oke, J.~B., Cohen, J.~G., Carr, M., Cromer, J., Dingizian, A., Harris, F.~H.,
             Labrecque, S., Lucinio, R., et al. 1995, PASP, 107, 375
\reference{} Ranalli, P., Comastri, A., \& Setti, G. 2003, A\&A, 399, 39
\reference{} Richards, E.~A. 2000, ApJ, 533, 611
\reference{} Seibert, M., Heckman, T.~M., Meurer, G.~R. 2002, AJ, 124, 46
\reference{} Stark, A.~A., Gammie, C.~F., Wilson, R.~W., Bally, J., Linke, R.~A., Heiles,
             C., \& Hurwitz, M. 1992, ApJS, 79, 77
\reference{} Steidel, C.~C., Adelberger, K.~L., Giavalisco, M., Dickinson, M., \&
             Pettini, M. 1999, ApJ, 519, 1
\reference{} Steidel, C.~C., Shapley, A.~E., Pettini, M., Adelberger, K.~L., Erb, D.~K.,
             Reddy, N.~A., \& Hunt, M.~P. 2004, submitted to ApJ
\reference{} Webb, T.~M., Eales, S., Foucaud, S., Lilly, S.~J., McCracken, H., Adelberger, K.,
             Steidel, C., Shapley, A., et.~al. 2003, ApJ, 582, 6
\reference{} Weedman, D., Charmandaris, V., \& Zezas, A. 2004, ApJ, 600, 106
\reference{} Yun, M.~S., Reddy, N.~A., \& Condon, J.~J. 2001, ApJ, 554, 803

\end{references}
\end{document}